\documentclass{llncs}
\usepackage[T1]{fontenc}
\usepackage{graphicx}
\usepackage{booktabs}
\usepackage{amsmath,amssymb}
\usepackage{xcolor}
\usepackage{enumitem}
\usepackage[hidelinks]{hyperref}

\newif\ifdraftnotes
\draftnotesfalse      

\ifdraftnotes
  \newcommand{\todo}[1]{\textcolor{red}{\textbf{[TODO: #1]}}}
  \newcommand{\needsrun}[1]{\textcolor{orange}{\textbf{[NEEDS EXPERIMENT: #1]}}}
  \renewcommand{\note}[1]{\textcolor{blue}{[#1]}}
\else
  \newcommand{\todo}[1]{}
  \newcommand{\needsrun}[1]{}
  \renewcommand{\note}[1]{}
\fi

\makeatletter
\renewcommand\paragraph{\@startsection{paragraph}{4}{\z@}%
                       {-8\p@ \@plus -2\p@ \@minus -3\p@}%
                       {-0.5em}%
                       {\normalfont\normalsize\itshape}}
\makeatother

\begin{document}

\title{Decompose, Estimate, Select: Coverage-Aware Context Construction for Long-Form RAG}
\titlerunning{Coverage-Aware Context Construction for Long-Form RAG}


\author{He Tu\inst{1} \and
        Aleena Bosco Kurumthottam\inst{2} \and
        Torsten Suel\inst{2}}
\institute{
           $^1$New York University Shanghai \quad
           $^2$New York University}

\maketitle


\begin{abstract}
Retrieval-augmented generation (RAG) grounds a language model's output in
retrieved evidence, but for complex questions spanning multiple aspects of an
information need, relevance-based retrieval may return redundant evidence while
leaving some aspects unsupported. We propose a coverage-aware retrieval pipeline
that decomposes queries into
subquestions, ranks candidates with a coverage-trained bi-encoder, predicts a
document\,$\times$\,subquestion coverage matrix with a cross-encoder, and
selects documents by greedy maximum-marginal coverage. On TREC RAG 2024, our
approach improves $\alpha$-nDCG@10, which rewards newly covered nuggets and
discounts redundant evidence, from $0.625$ to $0.702$, approaching the
$0.718$ gold-nugget upper bound without sacrificing relevance. We then find
that ground-truth oracle retrieval, despite substantially higher retrieval
coverage, delivers no more nugget coverage to the answer than our estimated
selection. This suggests that answer completeness depends not only on what
evidence is retrieved, but also on how effectively the generator uses it. To
address this, we introduce a \emph{grounded self-feedback loop} in which the
model identifies underused evidence from its first answer and generates again
without gold supervision, raising $V_{\mathrm{strict}}$, the fraction of vital
reference nuggets fully supported by the answer, by $0.086$ over two passes
while largely preserving faithfulness.

\end{abstract}


\section{Introduction}
\label{sec:intro}
Information retrieval has long centered on relevance: can a system
find documents relevant to a user query? Decades of progress, from statistical
ranking to neural retrievers, have been measured with graded-relevance metrics
such as nDCG and MAP. Retrieval-augmented generation (RAG) inherits this
framework but changes the role of retrieved documents: rather than being
returned to a user, they serve as evidence for downstream generation. In
long-form settings \cite{mayfield2024reports,rosset2024researchy,malaviya2024expertqa},
they therefore need to complement one another, not merely be individually
relevant.

A set assembled from independent relevance scores is usually highly redundant:
because documents are scored independently, ten documents elaborating the same
facet can score as well as ten that divide the query among them, although only
the latter provides evidence across all the nuggets a complete answer requires. On
TREC RAG 2024, a selector with gold document--nugget labels reaches $0.858$
subtopic recall (\textsc{StRecall}) with five documents, the same value it
attains at ten, whereas a strong coverage-trained retriever reaches $0.762$ with
ten. Relevance-based retrieval can therefore spend much of the context budget
on redundant evidence rather than expanding coverage.

Existing approaches address this in two ways, neither of which captures the
selection objective. Classical diversification penalizes documents resembling
those already selected, yet nugget-level redundancy need not correspond to
embedding similarity: a document can be relevant and dissimilar while still
contributing little new information. Training the retriever for coverage helps
\cite{cover2026}, but a bi-encoder scores documents independently and cannot
optimize a set-level objective.

We therefore build a retrieval pipeline that makes coverage explicit. It
decomposes each query into subquestions with an LLM, ranks candidates from a
first-stage pool with a coverage-trained bi-encoder, predicts a
document\,$\times$\,subquestion coverage matrix with a cross-encoder, and
selects documents by greedy maximum-marginal coverage. Both models are trained
on document--subquestion pairs from the Researchy Questions dataset
\cite{rosset2024researchy}, using LLM answerability judgments and the contrastive
coverage objective of CoveR \cite{cover2026}. On TREC RAG 2024 the pipeline
raises $\alpha$-nDCG@10 from $0.625$ to $0.702$, approaching the
$0.718$ bound obtained when reference nuggets replace generated subquestions,
without sacrificing relevance.

Improving retrieval coverage raises a further question: how much of the retrieved evidence reaches the answer? Holding the generator, judge, candidate pool, and
context budget fixed while varying only document selection, we find that the
relationship is not monotone. A gold-oracle selector covers substantially more
nuggets in the retrieved set, yet delivers no more of them to the answer than our
estimated selection. We address this with a grounded self-feedback loop, in
which the model identifies underused evidence from its first answer and
regenerates without gold supervision or additional retrieval, raising strict
vital-nugget coverage ($V_{\mathrm{strict}}$) by $0.086$ over two passes while
largely preserving faithfulness.

\paragraph{Contributions.} Our main contributions are as follows:
\begin{itemize}[itemsep=0pt,parsep=0pt,topsep=3pt,partopsep=0pt,leftmargin=1.4em]
\item A coverage-aware retrieval pipeline that replaces embedding-geometry
proxies for novelty with an explicitly predicted coverage matrix and greedy
set-level selection.
\item A controlled downstream study showing that higher retrieval coverage need
not yield higher answer coverage.
\item A \emph{grounded self-feedback loop} that improves answer coverage without
gold supervision or additional retrieval.
\end{itemize}

\noindent In the remainder of this paper, we review related work
(Section~\ref{sec:related}), describe our data and evaluation metrics
(Section~\ref{sec:data}), present the experimental framework for the retrieval
and generation settings (Section~\ref{sec:setup}), report our main results
(Section~\ref{sec:results}), and discuss their implications
(Section~\ref{sec:discussion}). 
\section{Related Work}
\label{sec:related}
In this section, we provide background and discuss the most relevant work.

\subsection{Multi-faceted Questions and Query Decomposition}
We study multi-faceted information needs, where a complete answer must synthesize distinct facts across several facets of a query. This is common in report-style requests \cite{mayfield2024reports,pradeep2024ragnarok}, decompositional web research questions \cite{rosset2024researchy}, and expert-written questions requiring attributed long-form answers \cite{malaviya2024expertqa}. Unlike questions centered on a single ambiguity \cite{min2020ambigqa,stelmakh2022asqa} or with multiple valid answers \cite{amouyal2023qampari}, these require one response covering many nuggets. Query decomposition is a common strategy \cite{chan2024rqrag,press2023selfask,salemi2025planrefine}, but decomposition alone does not determine which documents to retain across subquestions, whose candidate sets may overlap substantially.

\subsection{Information Coverage}
Selecting documents for what they \emph{add} to those already chosen is a long-standing idea in information retrieval, from subtopic retrieval \cite{zhai2003subtopic} and maximal marginal relevance \cite{carbonell1998mmr} to $\alpha$-nDCG \cite{clarke2008alphandcg} and aspect-based diversification \cite{chapelle2011intent,santos2010xquad}, with some recent neural approaches in \cite{yu2023aspectbottleneck,chen2024ma4div}. Long-form RAG extends this setting to queries with many subquestions and a context-window budget, where a generator consumes the selected documents. CoveR \cite{cover2026} adapts the retrieval objective to this setting, while other approaches optimize selection through document-set diversity or redundancy penalties \cite{li2024smartrag,ko2025adapcr,guo2026adagate,khurshid2026bubbles}. We adopt the retrieval objective of \cite{cover2026} and instead focus on selection, predicting an explicit document\,$\times$\,subquestion coverage matrix rather than using embedding-based proxies for novelty.

\subsection{Nugget-Based Evaluation}
Nugget-based evaluation measures long-form answers by the atomic facts they convey \cite{mayfield2024reports}. AutoNuggetizer \cite{pradeep2024autonuggetizer,pradeep2025nuggetrecall} automates nugget construction and answer assessment, while FActScore \cite{min2023factscore} applies atomic-fact decomposition to factual precision and ICAT \cite{samarinas2025icat} evaluates coverage of diverse facts. CoverageBench \cite{samuel2026coveragebench} further provides coverage-annotated collections across tasks and domains. We adopt these established evaluation protocols and use their metrics to measure answer coverage and faithfulness.

\subsection{Context Utilization and Iterative Generation}
Retrieved evidence is not necessarily used uniformly by the generator: performance can depend on the position of information in a long context \cite{liu2024lost}, degrade as context grows \cite{levy2024sametask}, and vary non-monotonically with the amount of retrieved evidence \cite{jin2024longrag}. These findings motivate studying context utilization beyond retrieval quality. Iterative RAG systems typically address incomplete answers by retrieving additional evidence \cite{asai2024selfrag,jiang2023flare,trivedi2023ircot,yan2024crag} or revising the generated response \cite{madaan2023selfrefine,salemi2025planrefine}, while recent nugget-guided approaches combine generation with explicit coverage signals \cite{dietz2026crucible,lajewska2025ginger}. In contrast, our loop keeps the retrieved set fixed and asks the model to identify underused evidence before regenerating, isolating improvements due to context utilization.

\section{Data and Evaluation Metrics}
\label{sec:data}

\subsection{Training Data}
\label{sec:data-train}

Both trained components use subquestion answerability judgments rather than
relevance labels. We use the CRUX release of Researchy Questions
\cite{rosset2024researchy}, where each query is paired with GPT-4-generated
subquestions and a candidate document pool. A Llama judge rates each
document--subquestion pair from $0$ to $5$. Following \cite{cover2026}, we treat
ratings $\geq 4$ as coverage, yielding a reference coverage matrix from which
both training sets are derived.

For the bi-encoder, we sample a partially covered state $C$ and label documents
by the coverage they add: \emph{positive} if they cover at least one
subquestion outside $C$, \emph{redundant negative} if they cover only
subquestions in $C$, and \emph{irrelevant negative} otherwise. This yields
$1{,}929$ training examples over 847 queries. A larger and more selectively
labeled variant ($44{,}558$ examples over $20{,}000$ queries), with covered states
sampled along a simulated retrieval trajectory and positives restricted to
documents covering more than half of the still-uncovered subquestions,
performs \emph{worse} ($\alpha$-nDCG@10 $0.598$ versus $0.656$): at round zero,
the covered set is empty, so the stricter threshold discards most moderately
relevant documents as neutral. We therefore use the binary variant throughout.

The coverage estimator is trained on $850{,}000$ document--subquestion pairs
from $3{,}000$ queries, using the same ratings $\geq 4$ target. Approximately
$69\%$ of pairs have the majority label, which we use as the accuracy baseline.

\subsection{Evaluation Collections}
\label{sec:data-eval}

All evaluation queries are held out from training. We use two collections from
CoverageBench \cite{samuel2026coveragebench}: TREC RAG 2024 with $56$ queries
and CRUX-DUC04 with $50$ queries.

We exclude TREC Fair Ranking 2022 because its ``nuggets'' represent demographic
attributes of authorship. On TREC RAG 2024, the nuggets are human-written but
document--nugget links are LLM-judged (approximately $69\%$ precision and $90\%$
recall against human assessment); CRUX-DUC04 uses LLM-generated subquestions and
links. The same judge is used for every system we compare, on both collections.

\subsection{Metrics}
\label{sec:metrics}

\paragraph{Retrieval.}
We use $\alpha$-nDCG@10 \cite{clarke2008alphandcg} as the primary retrieval
metric: it credits a document for the nuggets it contributes and discounts
nuggets already covered higher in the ranking, so a ranking is rewarded for
breadth rather than for repeating the same nugget. \textsc{StRecall}@10, the
fraction of a query's distinct nuggets covered anywhere in the top ten, is the
coverage-only view that ignores rank, and P@10 and nDCG@10 are reported to
verify that coverage is not gained at the cost of relevance.\footnote{Table~\ref{tab:retrieval} averages over all $56$ TREC RAG 2024
queries; the downstream experiments use the $54$ for which at least one selected
document covers a reference nugget, the other two scoring zero for every system.
Every system therefore reads higher on the $54$-query basis by a factor of
$56/54$: \textsc{StRecall}@10 is $0.762$ against $0.791$ for the
coverage-trained retriever and $0.858$ against $0.888$ for oracle greedy.
Comparisons are always made within one basis.}

\paragraph{Generation.}
We follow the AutoNuggetizer protocol
\cite{pradeep2024autonuggetizer,pradeep2025nuggetrecall}, which labels each
reference nugget as supported, partially supported, or unsupported. We report
$V_{\mathrm{strict}}$, the fraction of vital nuggets fully supported,
\emph{Answer \textsc{StRecall}}, the fraction of reference nuggets conveyed,
\emph{Transfer rate}, the fraction of nuggets available in the retrieved
context that reach the answer, and \emph{Sentence precision}, the fraction of
factual sentences supported by cited documents. Llama-3.3-70B is used as the
judge throughout.



\section{Problem Definition and Experimental Setup}
\label{sec:setup}
In this section, we define the problem and describe our pipeline and setup.

\subsection{Problem Formulation}
\label{sec:formulation}

Following the nugget-based evaluation literature
\cite{mayfield2024reports,pradeep2024autonuggetizer}, a \emph{nugget} is an
atomic fact that a complete response to a query should contain; a reference
answer is decomposed into a set of them, and a system is credited for the
nuggets its output conveys. We say a
document \emph{covers} a nugget when it contains the information, and an answer
\emph{conveys} a nugget when it states it; the distinction matters because our
central question is how much of what the context covers the answer conveys.

Let $q$ be a query and let $\mathcal{N}(q) = \{n_1,\dots,n_m\}$ be all the distinct
nuggets a complete response should convey. Let $\mathcal{D}_q$ be a first-stage document
candidate pool and let $\mathrm{cov}(d,n) \in \{0,1\}$ indicate that document $d$
\emph{covers} nugget $n$. Under a context budget $k$, coverage-aware retrieval selects
$S \subseteq \mathcal{D}_q$ with $|S| = k$ maximizing
\begin{equation}
\label{eq:maxcov}
  f(S) \;=\; \bigl|\,\{\, n \in \mathcal{N}(q) \;:\; \exists\, d \in S,\
  \mathrm{cov}(d,n) = 1 \,\}\,\bigr| .
\end{equation}

Two properties of Eq.~\eqref{eq:maxcov} shape our design. First, $f$ is defined
over \emph{sets}, which independent bi-encoder scores $s(q,d)$ cannot capture,
motivating the explicit selection stage. Second, $f$ is monotone and submodular,
so greedy selection by marginal gain achieves a $(1-1/e)$-approximation of the
optimum.

\begin{figure}[tb]
  \centering
  \includegraphics[width=\textwidth]{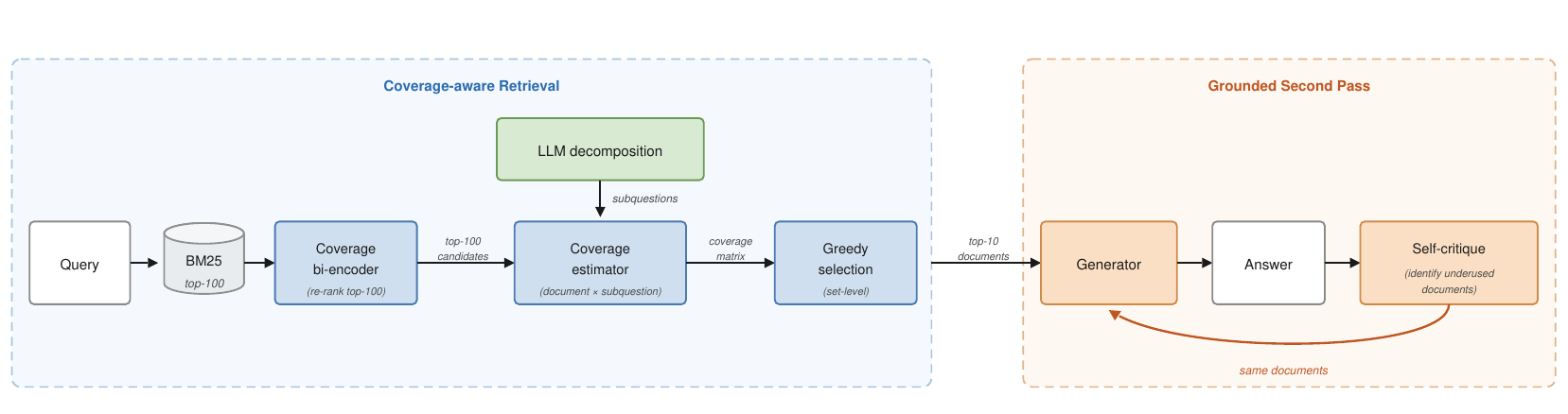}
  \caption{The coverage-aware pipeline. An LLM decomposes the query into
subquestions; a coverage estimator predicts a document\,$\times$\,subquestion
coverage matrix over the BM25 top-100 pool re-ranked by a coverage-trained
bi-encoder, and greedy maximum-marginal-coverage selection fixes a context of
$k = 10$. The grounded self-feedback loop re-generates from that
\emph{unchanged} context using no reference information.}
  \label{fig:pipeline}
\end{figure}

At inference time $\mathcal{N}(q)$ and $\mathrm{cov}$ are both unavailable. Our
pipeline estimates them: an LLM produces subquestions
$\tilde{\mathcal{N}}(q)$ in place of $\mathcal{N}(q)$, and a cross-encoder
predicts $\hat{C} \in [0,1]^{|\mathcal{D}_q| \times |\tilde{\mathcal{N}}(q)|}$ in
place of $\mathrm{cov}$ (Section~\ref{sec:framework-ret}). Figure~\ref{fig:pipeline} shows
the resulting system. Section~\ref{sec:results-ret} reports the cost of each
substitution by replacing it in turn with its reference counterpart.

\subsection{Retrieval Pipeline}
\label{sec:framework-ret}

\paragraph{First-Stage Pool.}
Every system in this paper re-ranks the same BM25 top-100 pool, so differences
reflect selection rather than first-stage recall.

\paragraph{Coverage-Trained Bi-Encoder.}
We initialize from a relevance-trained ModernBERT bi-encoder and continue
training on the pairs of Section~\ref{sec:data-train} with an InfoNCE objective (temperature $0.02$), using a shared encoder with \texttt{search\_query:} and
\texttt{search\_document:} prefixes, mean pooling, and $L_2$ normalization. The
anchor is the query paired with its simulated covered state, and the target is
one of its positive documents. The denominator contains two sources of negatives. The first is $n_{\text{neg}}$ documents
sampled from the query's own redundant and irrelevant buckets, with redundant
negatives included first so that the hardest negatives are always present. The
second is in-batch negatives: every other training example in the batch
contributes its own positive and its own sampled negatives, and all of them are
treated as negatives for the current anchor, so a batch of $b$ examples supplies
$(b-1)(n_{\text{neg}}+1)$ additional negatives per anchor at no extra encoding cost. Redundant and irrelevant negatives are not distinguished in
the objective, so the bi-encoder learns whether a document adds coverage rather
than why it fails to. This stage reproduces the training objective of
\cite{cover2026} on public data and is not a contribution of our work; we
include it so that the selection stage is measured on top of a competitive
coverage retriever rather than a relevance-only one.

\paragraph{Subquestion Decomposition.}
\label{sec:framework-decompose}
We prompt an LLM for ten self-contained subquestions per query, each wrapped

in \texttt{<q>} tags, following \cite{cover2026}.\footnote{\emph{Write 10 diverse
subquestions that can reveal the information required to answer the given report
query. Each subquestion should be self-contained and include the necessary
context. Collectively, the subquestions should fully cover the scope of the
report query.}}

Holding all other components fixed, we compare three decomposers: Qwen2.5-7B-Instruct, Llama-3.3-70B-Instruct, and GPT-4o. We also evaluate a \emph{document-grounded} variant that derives subquestions from the top-ranked pool documents rather than from the query alone.

\paragraph{Coverage Estimator.}
\label{sec:framework-ce}
A ModernBERT cross-encoder scores each (document, subquestion) pair, trained on
the distilled judgments of Section~\ref{sec:data-train} with a maximum input
length of $1024$ tokens. Applied to the pool it yields the matrix $\hat{C}$, which we binarize at
threshold $\tau$. We use $\tau = 0.3$ throughout unless otherwise stated;
Section~\ref{sec:results-ret} reports sensitivity to this threshold.

\paragraph{Greedy Selection.}
Starting from $S = \emptyset$ and with the full subquestion set as uncovered, we
repeatedly add the document covering the most still-uncovered subquestions at
threshold $\tau$, breaking ties by bi-encoder score, until $|S| = k$. Once no
candidate covers an uncovered subquestion the rule reduces to bi-encoder order,
so the output is always a complete ranking.

\paragraph{Reference Conditions.}
Two conditions isolate where the pipeline loses ground to the formulation of
Eq.~\eqref{eq:maxcov}. \emph{Gold nuggets} replaces the generated subquestions
with the reference nugget text but still predicts coverage with the estimator,
and therefore bounds what a perfect decomposer would buy. \emph{Oracle greedy}
replaces $\hat{C}$ with the reference document-nugget matrix, and therefore
bounds what any selection method can achieve on this pool.

\subsection{Generation Pipeline}
\label{sec:framework-gen}
\paragraph{Controlled Comparison.}
The following three conditions feed the \emph{same} generator from the \emph{same} pool at the
\emph{same} budget $k = 10$, assessed by the \emph{same} judge. Only document
selection varies:
\begin{itemize}[itemsep=0pt,parsep=0pt,topsep=3pt,partopsep=0pt,leftmargin=1.4em,label=$\bullet$]
  \item \textbf{A} Bi-encoder top-$k$, using relevance-ordered selection.
  \item \textbf{B} Our pipeline, using decomposition, coverage estimation, and greedy selection.
  \item \textbf{O} Greedy selection using the reference document-nugget matrix, providing an upper bound on retrieval-based selection.
\end{itemize}
Document text is truncated to $250$ words per document in all three conditions,
so context length does not confound the comparison.

\paragraph{Generation.}
The generator is Qwen2.5-7B-Instruct with greedy decoding and a $700$-token limit.
The prompt asks for a comprehensive answer in plain prose covering as many
distinct aspects as the documents support, consisting of about twelve sentences, with every
factual sentence ending in a citation to the documents it came from, and
forbids outside knowledge.

\paragraph{The Grounded Self-Feedback Loop.}
The second pass re-runs the generator over the \emph{unchanged} context of
condition B, asking for about eighteen sentences rather than the twelve
requested in the first pass, under the same citation requirement. Length alone
is not what produces the gain: asking the first pass for eighteen sentences
instead of twelve leaves transfer rate unchanged ($+0.000$) and answer
\textsc{StRecall} marginally lower ($-0.003$). We compare how the coverage gap
is identified:

\begin{itemize}[itemsep=0pt,parsep=0pt,topsep=3pt,partopsep=0pt,leftmargin=1.4em,label=$\bullet$]
  \item \textbf{iter2} injects the text of reference nuggets the first answer
        missed. This uses the evaluation key and is reported only as a leakage
        upper bound, never as a method.
  \item \textbf{iter2g} identifies the missed reference nuggets, maps them to the
        in-context documents that support them, and points the generator at those
        document identifiers without revealing nugget text.
  \item \textbf{iter2self} asks the generator to critique its own first answer
        and name the documents it under-used, then points it at those documents.
        This uses no reference information at any step and is the deployable
        configuration.
\end{itemize}
The contrast between \textbf{iter2} and the latter two isolates a design question:
whether feedback should \emph{supply} the missing content or \emph{redirect
attention} to evidence already present.

\section{Main Results}
\label{sec:results}

We present the main experimental results, first for retrieval and then for generation.
\subsection{Retrieval Results}
\label{sec:results-ret}

\begin{table}[tb]
\caption{Retrieval results on TREC RAG 2024 ($56$ queries, BM25 top-100 pool,
$k = 10$). Each row adds one component to the row above. We use
$\tau = 0.3$; \emph{gold nuggets} uses reference nugget text in place of generated
subquestions and is evaluated across the same threshold range reported in
Table~\ref{tab:decomposer}. \emph{Oracle greedy} replaces the predicted
coverage matrix with the reference document–nugget matrix, without threshold.}
\label{tab:retrieval}
\centering\footnotesize\setlength{\tabcolsep}{3.5pt}
\begin{tabular}{lcccc}
\toprule
Method & $\alpha$-nDCG@10 & \textsc{StRecall}@10 & P@10 & nDCG@10 \\
\midrule
Relevance-trained bi-encoder
    & $0.625$ & $0.744$ & $0.754$ & $0.784$ \\
\quad $+$ coverage training
    & $0.656$ & $0.762$ & $0.773$ & $0.798$ \\
\quad $+$ coverage estimator and greedy selection
    & $\mathbf{0.702}$ & $0.770$ & $0.775$ & $0.800$ \\
\midrule
\emph{Gold nuggets (decomposition upper bound)}
    & $0.718$ & $0.781$ & $0.771$ & $0.802$ \\
\emph{Oracle greedy (selection upper bound)}
    & $0.887$ & $0.858$ & $0.800$ & $0.857$ \\
\bottomrule
\end{tabular}
\end{table}

Table~\ref{tab:retrieval} decomposes the gain. Coverage training the bi-encoder on the pairs of
Section~\ref{sec:data-train} raises $\alpha$-nDCG@10 from $0.625$ to $0.656$, and the
explicit selection stage adds a further $0.046$ to reach $0.702$. The second step is larger, suggesting that with the retriever held fixed, modeling coverage as a set-level selection problem contributes more than modeling it within the retriever's training objective.

Two further properties of the gain matter as much as its size. First, it does not come from relevance: P@10 and nDCG@10 are flat or slightly higher at every
step, so the selection stage is not trading away precision for diversity.
Second, the two upper bounds separate cleanly. Replacing our generated
subquestions with reference nugget text recovers $0.016$ at its own best threshold, whereas replacing the
predicted coverage matrix with the reference one recovers a further $0.169$. At this operating point, the remaining headroom lies primarily in coverage \emph{estimation}, namely how accurately a document can be judged to answer a given nugget.

\begin{table}[tb]
\caption{$\alpha$-nDCG@10 on TREC RAG 2024 by decomposer and coverage threshold
$\tau$. All rows share the same bi-encoder, estimator and selection rule.}
\label{tab:decomposer}
\centering\footnotesize
\begin{tabular}{lccc}
\toprule
Subquestion source & $\tau = 0.3$ & $\tau = 0.5$ & $\tau = 0.7$ \\
\midrule
Qwen2.5-7B-Instruct
    & $0.675$ & $0.666$ & $0.663$ \\
Llama-3.3-70B-Instruct
    & $0.683$ & $0.669$ & $0.668$ \\
GPT-4o
    & $\mathbf{0.702}$ & $0.677$ & $0.672$ \\
GPT-4o, document-grounded
    & $0.672$ & $0.669$ & $0.653$ \\
\midrule
\emph{Gold nuggets (upper bound)}
    & $0.675$ & $0.690$ & $\mathbf{0.718}$ \\
\bottomrule
\end{tabular}
\end{table}

\paragraph{Decomposer Quality.}
Table~\ref{tab:decomposer} varies only the source of the subquestions. GPT-4o
scores highest at every threshold, followed by Llama-3.3-70B and Qwen2.5-7B, but
the returns diminish sharply: moving from a 7B to a frontier decomposer is worth
$0.027$, while the remaining distance to reference nuggets is $0.016$. Deriving
subquestions from retrieved documents instead of from the query is worse at every
threshold ($0.672$ versus $0.702$ at $\tau = 0.3$), because grounding inherits
the pool's errors and can only surface subquestions that some pooled document
already discusses. The reference row also prefers a strict threshold ($0.718$ at
$\tau = 0.7$) where generated subquestions prefer a loose one: reference nuggets
are precise and largely non-overlapping, whereas generated subquestions are
broader and partially redundant, so a strict criterion leaves too many of them
uncovered and the selection rule degenerates toward bi-encoder order.

\subsection{Generation Results}
\label{sec:results-gen}

\begin{table}[tb]
\caption{Downstream results on TREC RAG 2024. A, B and O share the same
generator, judge, candidate pool and context budget, differing only in document
selection; means are over the queries scorable under all three, with $n$ that
count. $\Delta$ is the paired mean difference B$-$O (two-sided paired $t$-test,
uncorrected); $^{\dagger}p<.05$, $^{\ddagger}p<.001$. None survives
Holm--Bonferroni across the five metrics.}
\label{tab:generation}
\centering\footnotesize
\begin{tabular}{lrcccr}
\toprule
 & & A & B & O & \\
Metric & $n$ & bi-encoder & ours & oracle & $\Delta$(B$-$O) \\
\midrule
\textsc{StRecall}@10 (retrieval)
    & $54$ & $0.791$ & $0.799$ & $\mathbf{0.888}$ & $-0.089^{\ddagger}$ \\
\midrule
Transfer rate
    & $54$ & $0.294$ & $\mathbf{0.326}$ & $0.275$ & $+0.050^{\dagger}$ \\
Answer \textsc{StRecall}
    & $54$ & $0.239$ & $\mathbf{0.266}$ & $0.247$ & $+0.020$ \\
$V_{\mathrm{strict}}$
    & $51$ & $0.350$ & $\mathbf{0.351}$ & $0.322$ & $+0.029$ \\
Sentence precision
    & $53$ & $\mathbf{0.715}$ & $0.711$ & $0.642$ & $+0.070^{\dagger}$ \\
\bottomrule
\end{tabular}
\end{table}

\paragraph{Higher Retrieval Coverage Need Not Yield Higher Answer Coverage.}
Table~\ref{tab:generation} separates what retrieval covers from what the answer
conveys. The oracle retrieves substantially more coverage than ours ($0.888$
vs.\ $0.799$), yet does not convey more of it to the answer: our selection is
ahead in transfer rate ($+0.050$, $p=.042$) and sentence precision ($+0.070$,
$p=.030$), and ahead but not significantly on answer \textsc{StRecall}
($+0.020$, $p=.32$) and $V_{\mathrm{strict}}$ ($+0.029$, $p=.42$); after
Holm--Bonferroni correction across the five comparisons none remains
significant. We read this as evidence that a large advantage in retrieved
coverage need not produce a more complete answer, not that our selection is
superior.
\paragraph{Coverage-Aware Selection Does Not Consistently Improve Answer Coverage.}
Compared with A, our selection gains $+0.032$ in transfer rate and $+0.027$ in
answer \textsc{StRecall} on TREC RAG 2024, neither significant ($p = .22$ and
$p = .10$). On CRUX-DUC04 the direction reverses: B retrieves slightly more
coverage than A (\textsc{StRecall}@10 $0.675$ versus $0.668$) but conveys less of
it, with transfer rate $-0.037$ ($p = .064$) and answer \textsc{StRecall}
$-0.024$ ($p = .041$). Pooled over both collections the difference in answer
\textsc{StRecall} is $+0.003$ ($p = .78$). We therefore do not claim that
coverage-aware selection consistently improves downstream coverage over
relevance-ordered selection. $V_{\mathrm{strict}}$ is not reported across
collections because CRUX nuggets carry no importance labels.

\begin{table}[tb]
\caption{Successive passes of the grounded self-feedback loop on TREC RAG 2024.
Every column uses the ten documents selected by condition B of
Table~\ref{tab:generation}: B is that condition's single answer, and $+k$ passes
adds $k$ further generations over those same documents, each preceded by a
self-critique. $\Delta$ compares $+$2 passes against B (two-sided paired
$t$-test); $^{\dagger}$ survives Holm--Bonferroni across the five metrics. Each
table pairs over its own conditions, so B's means differ slightly from
Table~\ref{tab:generation}.}
\label{tab:rounds}
\centering\footnotesize
\begin{tabular}{lrccccr}
\toprule
Metric & $n$ & B & $+$1 pass & $+$2 passes & $+$3 passes & $\Delta$ \\
\midrule
\textsc{StRecall}@10 (retrieval)
    & $54$ & $0.799$ & $0.799$ & $0.799$ & $0.799$ & $+0.000$ \\
\midrule
Transfer rate
    & $54$ & $0.326$ & $0.376$ & $\mathbf{0.391}$ & $0.363$ & $+0.065^{\dagger}$ \\
Answer \textsc{StRecall}
    & $54$ & $0.266$ & $0.289$ & $\mathbf{0.308}$ & $0.295$ & $+0.042^{\dagger}$ \\
$V_{\mathrm{strict}}$
    & $53$ & $0.338$ & $0.406$ & $\mathbf{0.424}$ & $0.421$ & $+0.086^{\dagger}$ \\
Sentence precision
    & $54$ & $\mathbf{0.703}$ & $0.652$ & $0.661$ & $0.655$ & $-0.042$ \\
\bottomrule
\end{tabular}
\end{table}

\paragraph{The Grounded Self-Feedback Loop Reaches the Answer.}
Table~\ref{tab:rounds} applies the loop to condition B's context, so retrieval is
identical by construction and all changes are attributable to context use.
Coverage improves for two passes and then plateaus: $V_{\mathrm{strict}}$ gains
$+0.068$ after the first pass and a further $+0.018$ after the second, reaching
$+0.086$ over B, with transfer rate and answer \textsc{StRecall} following the
same pattern. After two additional passes, three of the five measures remain
significant under Holm--Bonferroni correction across that column (answer
\textsc{StRecall} adjusted $p = .023$, $V_{\mathrm{strict}}$ $.029$, and
transfer rate $.036$). When correcting instead across the three round
comparisons for $V_{\mathrm{strict}}$, our pre-declared primary metric, only the
two-pass configuration remains significant ($p = .022$ versus $.059$ for one
pass). A third pass reverses part of the gain (transfer rate $-0.027$, answer
\textsc{StRecall} $-0.013$, $V_{\mathrm{strict}}$ $-0.002$) and yields no
significant result after correction. We therefore treat two passes as an
empirical stopping point rather than a budget constraint, although each pass
requires an additional generation over the full context.

Sentence precision drops by $0.050$ after the first pass and then remains stable,
ending $0.042$ below B after two passes and $0.048$ below B after three
($p = .210$ and $.144$). Thus, the precision loss occurs at the first
re-generation and does not accumulate with further passes. A combined $F_1$ of
answer \textsc{StRecall} and sentence precision follows the same pattern,
peaking after two passes ($+0.033$, uncorrected $p = .037$). Two passes
therefore improve coverage without a net faithfulness cost, whereas the third
retains the precision loss without the coverage gain.

\paragraph{Feedback Should Point at Evidence, Not Supply It.}
Replacing the self-critique with the text of the missed reference nuggets
(\textbf{iter2}) drives transfer rate to $0.771$, but sentence precision falls
from $0.703$ to $0.413$: the generator restates supplied text it cannot
attribute to any document. Pointing instead at the in-context documents that
support those nuggets (\textbf{iter2g}) keeps faithfulness intact and still
gains $+0.083$ in $V_{\mathrm{strict}}$ over B, with sentence precision down
only $0.031$ ($p = .28$). The model's own critique (\textbf{iter2self}) recovers
about $80\%$ of that grounded gain ($+0.068$) using no reference information at
any step. The distance between iter2's $0.771$ and iter2g's $0.383$ transfer
rate is the part of the coverage gap that redirecting attention cannot close; we
therefore report \textbf{iter2} only as a leakage bound.

\section{Discussion}
\label{sec:discussion}

We now discuss the main conclusions from our work and its limitations.

\paragraph{More Retrieval Coverage May Not Improve Answers.}
The oracle selector (O) increases retrieved coverage substantially and
significantly, yet none of the four downstream measures improves: additional
evidence is not automatically useful once it enters the context. Improving the
coverage estimator may therefore raise retrieval metrics without a corresponding
downstream gain.

\paragraph{Context Utilization As Bottleneck.}
The oracle result suggests that maximizing the diversity of evidence is not
necessarily the right downstream objective. A context containing more distinct
nuggets will also tend to contain documents that share fewer facts with one
another, so within the same fixed answer length the answer must move between
unrelated points instead of developing a few of them. This is consistent with
prior evidence that language models use long and densely packed contexts
unevenly \cite{liu2024lost,levy2024sametask,jin2024longrag}, although our
experiments do not isolate this mechanism.

\paragraph{Generalization and Evaluation Limitations.}
Our collections are small ($56$ and $50$ queries). One pass of the loop
replicates on CRUX-DUC04 (transfer rate $+0.062$, $p = .006$; answer
\textsc{StRecall} $+0.039$, $p = .005$), and pooled across the two collections ($n = 104$) both gains survive Holm
correction ($+0.056$ and $+0.030$, both
$p = .007$). On CRUX-DUC04, the judge failed to parse $7$ of $50$ answers for
B and $13$ of $50$ after the loop, so the gains persist despite more failures.
The multi-pass schedule in Table~\ref{tab:rounds} was evaluated only on TREC
RAG 2024, and $V_{\mathrm{strict}}$ cannot be pooled because CRUX-DUC04 has no
nugget importance labels. Three further caveats are that $\tau$ is chosen on
the evaluation collection rather than a held-out set, so $\tau = 0.3$ is
optimistic; we ran no significance tests for Table~\ref{tab:retrieval}, so its
ordering is indicative only; and our best configuration uses a proprietary
decomposer, although the fully open Llama-3.3-70B-Instruct costs only $0.019$
in $\alpha$-nDCG@10 ($0.683$ versus $0.702$), making a fully open pipeline
available at modest cost. Finally, nugget links are LLM-judged on both
collections, so the shared judge keeps comparisons internally consistent but
absolute scores may inherit its biases. The bi-encoder is also trained on only
$1{,}929$ examples, and training-set size is not separated from the labeling
rule that makes the larger variant worse.


\section{Conclusion}
\label{sec:conclusion}
We now summarize our main conclusions.
Making coverage explicit improves retrieval. An estimated
document\,$\times$\,subquestion coverage matrix with greedy
maximum-marginal-coverage selection raises $\alpha$-nDCG@10 from $0.625$ to
$0.702$ on TREC RAG 2024 without sacrificing relevance, with more of the gain
coming from selection than from coverage training of the bi-encoder. Higher
retrieval coverage, however, does not translate directly into better answers.
Holding everything but selection fixed, a gold-oracle selector with
substantially higher retrieval coverage conveys fewer nuggets to the answer
than our estimated selection, with downstream transfer remaining near $0.30$
across the retrieval-coverage range we tested. Answer coverage instead
improves when the model is pointed back to underused evidence through a
grounded self-feedback loop over the same context. For long-form RAG under a
fixed context budget, these results favor improving evidence use over simply
increasing retrieval coverage. Two questions remain open. First, the remaining retrieval headroom lies
largely in coverage \emph{estimation}: whether it can be closed with scale or
requires supervision beyond distilled answerability judgments is unclear.
Second, if the generator limits answer completeness, selecting evidence for
how usable it is, rather than only for what it covers, is a natural next step.


\bibliographystyle{splncs04}
\bibliography{refs}

\end{document}